\documentclass{article}
\usepackage{epsfig}
\usepackage{graphicx}
\usepackage{natbib}
\title{Performance metrics}
\begin{document}

\maketitle
\noindent
It can be argued that the quality of science cannot be measured purely by
quantitative metrics, and certainly not by a single one that is
supposed to fit all. Experimental groups and theoretical groups have
different types of output and it is difficult to envisage a way in which 
the impact or quality of these can be compared either objectively or
subjectively. A single
instrument built by an experimental group may well lay the foundation
for very many scientific results and papers by opening up a new field,
and a single theoretical paper may well cause a paradigm shift. In
those branches of physics where close links with industrial
applications can be forged, yet another measure of succes would need
to be found.

Even for those research groups for which the primary scientific output
could be considered to be scientific papers, the rate at which papers 
are published, and therefore also are cited depends on the field. Even
within the area of astrophysics, in some subfields papers are published
at rather larger rates than in others. No-one could reasonably claim
that the former sub-fields have greater scientific significance or
importance than the latter. In this sense it seems that the
conclusions of Pearce (1994) on `good researchers' should not be used
without additional judgment, as indeed pointed out by that author.

\begin{figure}
\centerline{\psfig{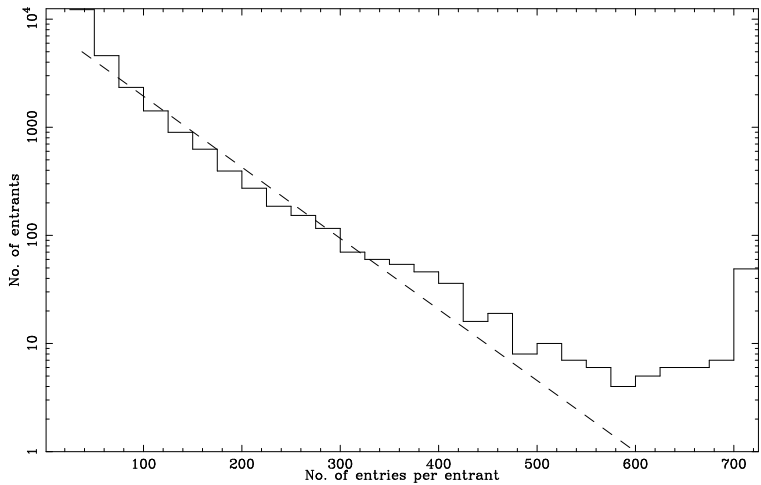}}
\vspace*{8pt}
\caption{Histogram of the number of bibliographic entries per author
in the ADS against the number of authors. The dashed line shows an
exponential distribution.}
\end{figure}
To illustrate this point I have compiled some citation statistics by
making use of the Smithsonian/NASA Astrophysics Data System (ADS)
 which also tracks citation statistics (the UK mirror
site can be found at http://ukads.nottingham.ac.uk/).
The number of bibliographic entries, both refereed and non-refereed, 
per author in the ADS covers a wide range. Evidently there are many
authors with one or a few entries, and at the other end of the scale
there is an author for which there are in excess of 11000 entries. 
A histogram showing the distribution of authors in the ADS is shown :
each bin shows the amount of authors, against the number of
bibliographic entries per author.

An author with very many entries is likely to be better known, but it
does not seem likely that the scientific impact of e.g. each of 11000+
entries is the same. The number of citations a paper receives by
itself does not necessarily reflect impact either. A paper that is
authored by many can easily get a high citation ranking if each of the
authors subsequently writes a further paper in which the original
paper is cited. If an average paper in a given field of research has a
certain a-priori likelihood of being cited, i.e. the standard impact
of a paper in that field, the number of citations it receives must
scale with the number of papers published in that field. Also, the
rate at which a typical author publishes will reflect the habits of
that subfield so that one should expect that for an unexceptional
author the number of citations scales with the number of
papers that this author publishes to some power that lies between 1
and 2. 
\begin{equation}
N_{\rm expected\ citations} \propto N_{\rm papers\ 
published}^\gamma\hskip 1cm 1<\gamma <2
\end{equation}
From the ADS it is possible to extract the normalised citation rate of
papers of any author, which is calculated by taking for each paper the
number of citations it has received, divides by the number of authors,
and then sums over all entries/papers on which this author appears.
This to some extent accounts for differences between many author and
single author papers.

I have randomly sampled, from the distribution shown in Fig. 1,
authors for whom I have subsequently extracted this normalised 
citation count for
those bibliographic entries which are refereed papers. The
sampling was carried out as a constant fraction $0.02$ per bin, rather than
for the distribution as a whole in order to reduce the statistical
fluctuations and cover the entire range reasonably evenly. Also,
authors with less than 50 bibliographic entries were omitted from the
analysis in order to reduce statistical fluctuations in citation counts.  
\begin{figure}
\centerline{\psfig{file=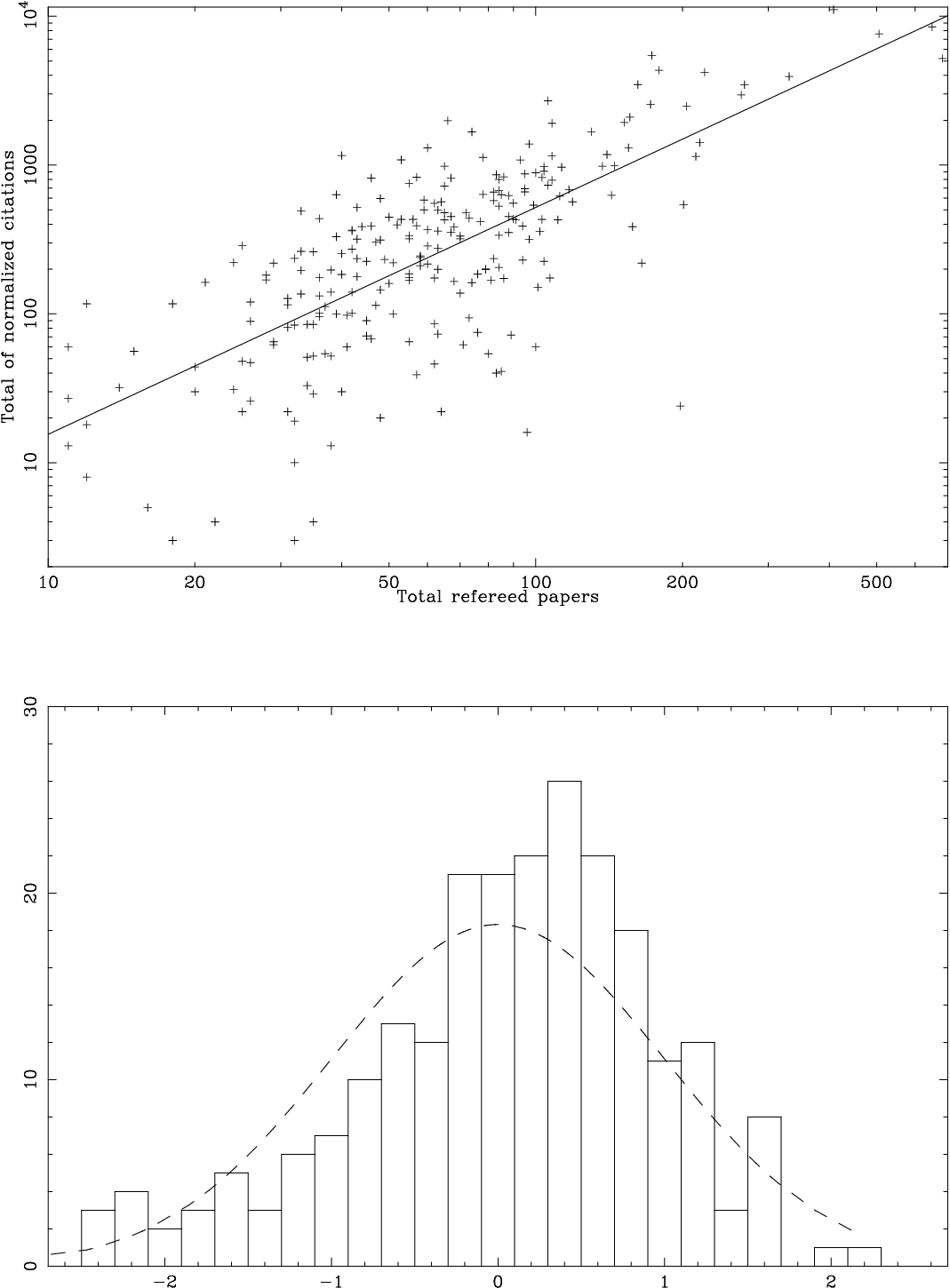,width=10.cm}}
\vspace*{8pt}
\caption{Upper panel: for each author the number of normalised citations summed
over all refereed papers is plotted against the number of refereed 
papers published by that author. Lower panel, the distribution around
the least squares fit shown as dashed line in the upper panel. }
\end{figure}
The random sampling within each bin sometimes produced a name for
which bibliographic entries could not uniquely be assigned. Whenever
this occured the name was discarded and a different, randomly selected,
name was substituted. 
The diagram resulting from this is shown on log-log scales in which
the solid line is the least squares fit with a $\gamma = 1.52$.
The behaviour appears to correspond to the expectation outlined above.
In absolute terms authors appearing in the upper right-hand corner
of this diagram, such as e.g. Prof. Sir Martin Rees, have great
visibility and therefore impact in astrophysics.
An equally useful measure of the relative importance of
individual authors or groups of authors is the distance above or below
the fitted function indicating a relatively high or low number of
citations given the number of papers published by that author. An
example of this is given by the strongest upward outlier to this distribution
that I have found which is Alan Guth, whose seminal work in inflation
theory clearly has a very wide impact in current cosmology.
His entry is not part of the random sample, and therefore also not 
present in Fig. 2 but corresponds to a $3-\sigma$ upwards
outlier.
The bottom panel shows the distribution orthogonal to the fitted
function, with positive offset corresponding to authors with
normalised citation rates above the fit, and negative offset to those
below that fit, with overplotted a Gaussian distribution. The distribution
is quite evidently skewed, with a long left-ward tail, and possibly
somewhat more peaked than a Gaussian distribution. 
\begin{eqnarray}
{\rm skewness} &=& -1.09 \pm 0.25 \\
{\rm kurtosis} &=&  1.4 \pm 0.6
\end{eqnarray}
Creating a distribution such as this for a department or the national
community requires having available a complete citation and
publication record. Even if this were to exist, as argued above, it 
does not measure reliably the performance of those groups for which
scientific papers is not the only or primary scientific output. Any
attempt at measuring performance solely in this manner is therefore 
in practice impossible and in principle flawed.

Moreover, a measure such as this covers the entire career of
individuals. It is known that exceptional papers distinguish
themselves primarily through being cited consistently over decades,
rather than passing through a brief peak of citations and then
disappearing from citation records. Even if the above were a reliable
measure it is not amenable to adaptation for the very short-term
measures sought for RAE assessment exercises and the like. To the
authors' knowledge there is no metric that reliably measures
performance over periods as short as 5 years, and none that is an
indicator of future performance. Specifically it seems inappropriate
to use numbers of citations, even after `normalisation' in the sense
of the ADS, as a reliable direct metric of impact. 

This research has made use of NASA's Astrophysics Data System
Bibliographic Services.

\medskip\noindent
Pearce, F., (1994), Astronomy \& Geophysics, {\bf 45}, 2.15
\end{document}